\documentclass[twocolumn,showpacs,preprintnumbers,amsmath,amssymb]{revtex4}

\usepackage{graphicx}
\usepackage{color}
\usepackage{dcolumn}
\usepackage{bm}

\begin{document}

\preprint{APS/123-QED}

\title{Effect of attractions on correlation length scales in a glass-forming liquid}

\author{Wen-Sheng Xu, Zhao-Yan Sun\footnote{Correspondence author. E-mail: zysun@ciac.jl.cn}, and Li-Jia An\footnote{Correspondence author. E-mail: ljan@ciac.jl.cn}}
\affiliation{State Key Laboratory of Polymer Physics and Chemistry,
Changchun Institute of Applied Chemistry, Chinese Academy of
Sciences, Changchun 130022, People's Republic of China}



\date{\today}

\begin{abstract}
There is growing evidence that slow dynamics and dynamic heterogeneity possess structural signatures in glass-forming liquids. However, even in the weakly frustrated glass-forming liquids, whether or not the dynamic heterogeneity has a structural origin is a matter of debate. Via molecular dynamics simulation, we present a study of examining the connection between dynamic heterogeneity and bond orientational order in a weakly frustrated glass-forming liquid in two dimensions by taking advantage of assessing the effect of attractions on the correlation length scales. We find that attractions can strongly affect relaxation dynamics, dynamic heterogeneity and the associated dynamic correlation length of the liquid, but their influence on bond orientational order and the associated static correlation length shows a manner reminiscent of the effect of attractions on the thermodynamics of liquids. This implies that the growth of bond orientational order and static correlation length scale might be merely a manifestation of favoring the configurational entropy in weakly frustrated glass-forming liquids. Thus, our results lead strong evidence that bond orientational order cannot provide a complete description of dynamic heterogeneity even in weakly frustrated glass-forming systems.
\end{abstract}


\maketitle

\section{Introduction}

Probing correlation length scales in amorphous systems has become the subject of much current activity due to their important role in understanding the physical mechanism of the glass transition~\cite{Berthier1, Mosayebi, Sausset1, Charbonneau}. Since the collective nature of the dynamics of supercooled liquids approaching the glass transition point has already indicated the existence of a growing dynamic length scale~\cite{Glotzer1, Glotzer2, DH}, a deeper understanding of the correlation length scales allows to explain the glassy phenomena such as slow dynamics and dynamic heterogeneity. For example, evidence has been provided by studying higher-order static correlations~\cite{Tanaka1, Tanaka2, Sausset2} or analyzing local geometric structures~\cite{Coslovich, Pedersen} that slow dynamics and dynamic heterogeneity possess some structural features in some model glass-formers, which sheds insight into the structural origin of slow dynamics in glass-forming liquids. On the other hand, the study of correlation length scales could facilitate the testing of theoretical predictions since most theories of the glass transition have treated characteristic length scales as a key ingredient~\cite{Berthier1, Tarjus, Biroli1}. Moreover, understanding of the glass transition has been improved recently by studying the point-to-set correlations due to their generic features in describing spatial information of glass-forming liquids~\cite{PTS1, PTS2, PTS3, PTS4}.

The growing dynamic length scale in glass-forming liquids has triggered a large body of work to search connections between dynamics and structure~\cite{WidmerCooper1,
WidmerCooper2, WidmerCooper3, WidmerCooper4}. In this situation, conventional methods such as the static pair correlations cannot be used to detect structural order since they do not show dramatic change on the approach to the glass transition point. Instead, the higher-order static correlations has been frequently emphasized within the glass community. An intriguing result, emerging in recent years, is that the bond orientational order has been suggested to be the origin of slow dynamics and dynamic heterogeneity in the so-called weakly frustrated glass-forming liquids~\cite{Tanaka1, Tanaka2}. By using polydispersity to control the strength of frustration, Kawasaki and coworkers~\cite{Tanaka1, Tanaka2} have shown that in a weakly frustrated glass-forming liquid there exist transient clusters of highly ordered particles, whose size and lifetime increase towards the glass transition point. They also found that particles in the clusters are less mobile than others and that both dynamic length (characterizing the increasing heterogeneity of the dynamics) and structural length (characterizing the spatial extension of the local bond orientational order) tend to diverge at the ideal glass transition point. Thus it may suggest a static thermodynamic origin of dynamic heterogeneity rather than a purely dynamic one at least for the weakly frustrated glass-forming liquids. However, there are also studies which show a decoupling between the structural correlation length and the dynamic one, e.g., Sausset and Tarjus~\cite{Sausset2} observed that the static correlation length of a weakly glass-forming liquid in negatively curved space first grows and then saturates as temperature decreases whereas the dynamic correlation length always increases with decreasing temperature. Therefore, even in the weakly frustrated glass-forming liquids, the origin of slow dynamics and dynamic heterogeneity is still unclear and whether the bond orientational order can provide a complete description of dynamic heterogeneity remains elusive.

In this work, we examine the connection between dynamic heterogeneity and bond orientational order in a weakly frustrated glass-forming liquid by taking advantage of assessing the effect of attractions on the correlation length scales. Our study is inspired by the results of effect of attractions on the structure and the dynamics of liquids, which has been reevaluated recently in the context of glass formation~\cite{Attract1, Attract2}. By comparing the static pair structure and the relaxation dynamics of a binary Lennard-Jones (denoted by LJ in the following) glass-former and its corresponding purely repulsive variant (proposed by Weeks, Chandler and Andersen~\cite{WCA} and denoted by WCA in the following), Berthier and Tarjus~\cite{Attract1, Attract2} have shown that attractions can affect the relaxation dynamics of liquids in both quantitative and qualitative ways but have little effect on the static pair correlations. From these results, one can also expect significant effect of attractions on dynamic heterogeneity of supercooled liquids since it has been well established that the dynamic slowing down accompanies the development of dynamic heterogeneity. Thus, a crucial test of the role of bond orientational order in weakly frustrated glass-forming liquids can be made by examining whether similar effect of attractions occurs on bond orientational order and the associated static correlation length or not.

We demonstrate via molecular dynamics simulation that attractions can strongly affect the dynamic heterogeneity of the liquid and the associated dynamic correlation length. However, their influence on the bond orientational order and the associated static correlation length shows a different manner and is reminiscent of the effect of attractions on the thermodynamics of liquids. In the supercooled regime, dependence of the dynamic correlation length on the structural relaxation time is nearly the same and shows a power-law relation for models with and without attractions. Although we can also identify a power-law relation between the static correlation length and the structural relaxation time in the same temperature range, the exponent of the power laws  for the two models is not the same. This implies that the growth of bond orientational order and static correlation length scale might be merely a manifestation of favoring the configurational entropy in weakly frustrated glass-forming liquids. Thus, our results lead strong evidence that the bond orientational order cannot provide a complete description of dynamic heterogeneity even in weakly frustrated glass-forming systems.

\section{Model and methods}

We compare the dynamical and structural properties of a LJ glass-forming liquid in two dimensions as well as its corresponding WCA variant. The pair potential for the two systems is given by
\begin{equation}
U_{jk}(r)=\left\{\!\!\!
\begin{array}{ll}
4\epsilon [(\sigma_{jk}/r)^{12}-(\sigma_{jk}/r)^{6}+C_{jk}], & \text{for $r<r_{jk}^{c}$}\\
0, &
\text{otherwise,}
\end{array}\right.
\end{equation}
where $\epsilon$ is the depth of the potential well, $r$ is the distance between two particles and $\sigma_{jk}=(\sigma_{j}+\sigma_{k})/2$ with $\sigma_{j}$ the diameter of particle $j$, $r_{jk}^{c}$ is equal to the position of the minimum of $U_{jk}(r)$ for the WCA model (i.e., $2^{1/6}\sigma_{jk}$) and to a conventional cutoff of $2.5\sigma_{jk}$ for the LJ model, and $C_{jk}$ is constant which is fixed such that $U_{jk}(r_{jk}^{c})=0$. The difference between the two potentials is that the LJ potential has a long-ranged attractive tail while the WCA potential has only one short-ranged repulsive component. To form a weakly frustrated glass-forming system, the particle diameters uniformly distribute in the range $0.8-1.2$ with an interval of $0.001$, then the size polydispersity for the system is $\Delta=\sqrt{(<\sigma^{2}>-<\sigma>^{2})}/<\sigma>=11.59\%$, where $<\cdot\cdot\cdot>$ is the average of the corresponding variable among all the particles. Such system prevents occurrence of crystallization and tends to form a local hexagonal structure, and one can readily identify the static structural order and detect the corresponding spatial correlations.

We employed molecular dynamics simulation for a system with the particle number $N=6400$ in the $NVT$ ensemble, where Newton's equations of motion are integrated with the velocity form of the Verlet algorithm under periodic boundary conditions and the temperature $T$ is maintained by the Nos\'{e}-Hoover thermostat~\cite{Frenkel1}. All the particles have the same mass $m$. Length, time and temperature are reported in units of $<\sigma>$, $\sqrt{m<\sigma>^{2}/\epsilon}$ and $\epsilon/k_{B}$ with $k_{B}$ the Boltzmann's constant. The time step is $\Delta t=0.005$. The number density is fixed at $\rho=N/L^{2}=0.95$ ($L\simeq82.08$) with $L$ the box dimension. We also performed simulations for a system with $N=1000$ ($L\simeq32.44$) in order to check the possible finite size effects and we found that similar effect of attractions also occurs for the smaller system. We note that the finite size effects do affect the results of dynamic correlation length in simulations, as pointed out recently~\cite{Sastry1, Sastry2, Szamel}. However, our aim is to correctly identify the effect of attractions on the correlation length scales, rather than to quantify a true liquid. Thus, we only present the results for the system with $N=6400$ in this paper. At each state point, the system was first equilibrated for at least $100\tau_{\alpha}$ ($\tau_{\alpha}$ is the structural relaxation time and see below for its definition) before collecting data. In order to obtain reliable results and improve the statistics, we performed a production run for at least $500\tau_{\alpha}$ and $8$ independent runs.

\section{Results and discussion}

In this section, we first compare the relaxation dynamics and the static pair correlations of LJ and WCA models. Then we discuss the effect of attractions on dynamic heterogeneity, bond orientational order and the associated correlation lengths.

\subsection{Relaxation dynamics and static pair structure}

\begin{figure}[tb]
 \centering
 \includegraphics[angle=0,width=0.45\textwidth]{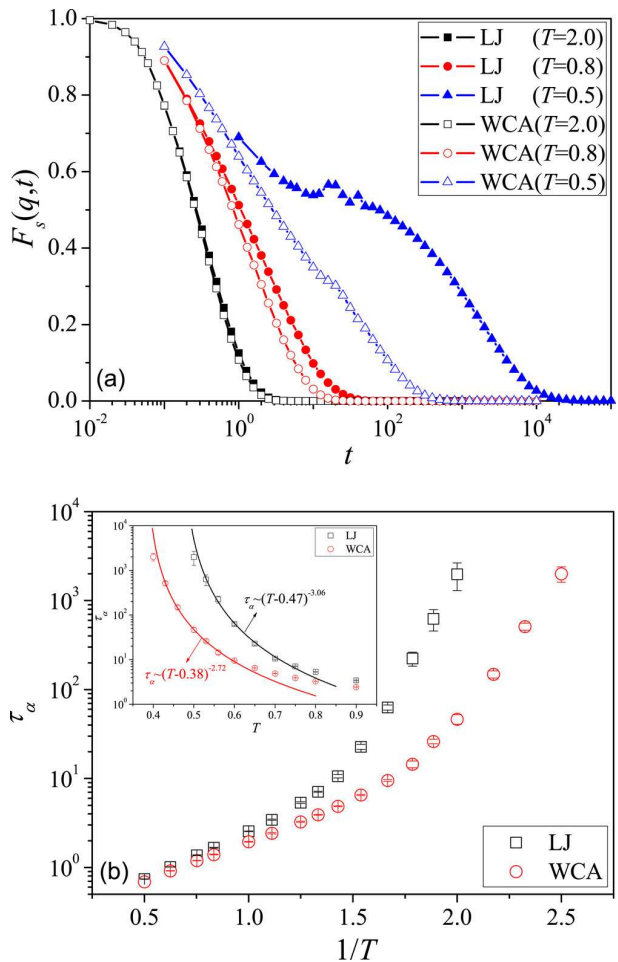}
 \caption{(a) The self-intermediate scattering function for LJ and WCA models at several temperatures. (b) Arrhenius plot of the structural relaxation time $\tau_{\alpha}$ for LJ and WCA models. Inset: Power-law fittings of $\tau_{\alpha}$ (see text). The fitting temperature ranges are $0.53-0.75$ and $0.43-0.6$ for LJ and WCA models respectively. The fitting results are: $T_{c}=0.47$, $\gamma=3.06$ for the LJ model and $T_{c}=0.38$, $\gamma=2.72$ for the WCA model.}
\end{figure}

We first consider the effect of attractions on the relaxation dynamics of liquids by calculating the self-intermediate scattering function (ISF), which is defined as
\begin{equation}
F_{s}(q,t)=\frac{1}{N}<\sum_{j=1}^{N}\exp\{i\textbf{q}\cdot[\textbf{r}_{j}(t)-\textbf{r}_{j}(0)]\}>,
\end{equation}
where $<\cdot\cdot\cdot>$ indicates the thermal average, $i=\sqrt{-1}$ and the wave number $q\simeq6.5$ corresponds to the first peak of the static structure factor (which will be shown later). The typical results are shown in Fig. 1(a). Consistent with the results in a binary glass-former~\cite{Attract1, Attract2}, we observe that although the difference of $F_{s}(q,t)$ between the two systems is very small at high temperatures, it becomes significant as the liquid enters the supercooled regime where typical two-step relaxation process occurs. This phenomenon can be more clearly seen from the dependence of the structural relaxation time $\tau_{\alpha}$ on $T$, which is shown in Fig. 1(b). Here, we define $\tau_{\alpha}$ as $F_{s}(q,t=\tau_{\alpha})=0.2$ (we have checked that choosing other reasonable values will not alter the qualitative results). We find that $\tau_{\alpha}$ for the two models is almost identical at $T=2.0$, but $\tau_{\alpha}$ of the LJ model is over $2$ orders of magnitude larger than that of the WCA model at $T=0.5$. We can also confirm the quantitative effect of attractions by comparing the mode-coupling glass transition point $T_{c}$ of the two models, as determined by power-law fittings of $\tau_{\alpha}$ (i.e., $\tau_{\alpha} \sim (T-T_{c})^{-\gamma}$)~\cite{MCT}. The results are presented in the inset of Fig. 1(b) and we find that $T_{c}^{LJ}$ is significantly larger than $T_{c}^{WCA}$ for the studied density (note that the difference between them will become smaller at higher densities~\cite{Attract1, Attract2}). As demonstrated in Refs.~\cite{Attract1, Attract2}, the effect of attractions on the dynamics is not only quantitative but also qualitative. The latter was evidenced by the absence in the WCA model of the density scaling of $\tau_{\alpha}$ which holds in the LJ model. We have also calculated the relaxation times at various number densities (ranging from $0.85$ to $1.05$) for the system with $N=1000$ and found the distinct qualitative behavior for the two models. Thus, the quantitative and qualitative effects of attractions on the relaxation dynamics are independent of dimensionality and should be universal in glass-forming liquids.

\begin{figure}[tb]
 \centering
 \includegraphics[angle=0,width=0.45\textwidth]{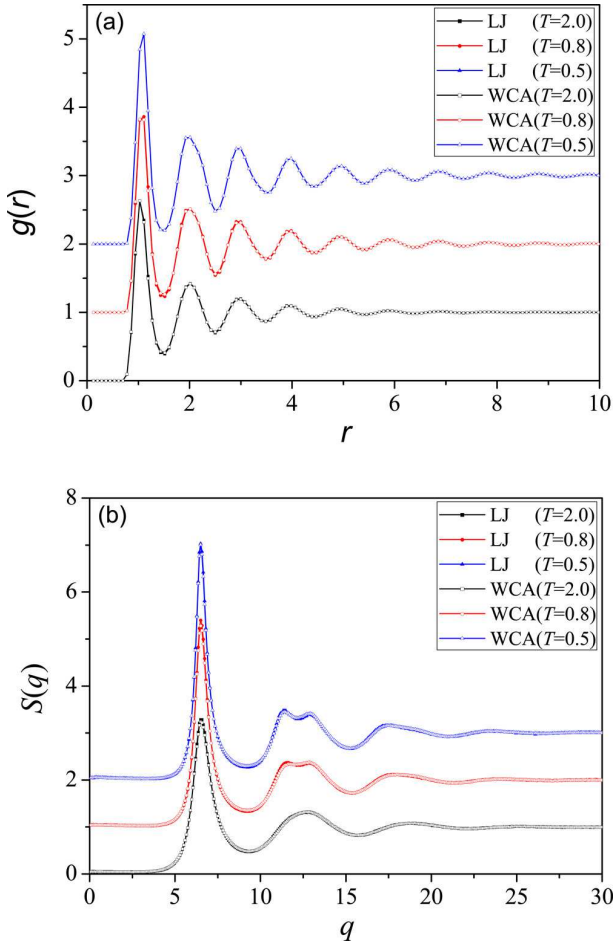}
 \caption{(a) Pair correlation function and (b) static structure factor for LJ and WCA models at several temperatures. The results have been shifted for clarity. Note that the solid and open symbols are indistinguishable due to the almost perfect collapse of data for the two models.}
\end{figure}

We present the results of the static pair correlations for the two systems at several temperatures in Fig. 2. Clearly, attractions have negligible effects on the pair correlation function $g(r)$ and the static structure factor $S(q)$ even in the supercooled regime. It should be noted that although the long-range positional order is obviously prevented in the two systems, the splitting second peaks in both $g(r)$ and $S(q)$ become more apparent as $T$ decreases (note that this phenomenon is more evident for $S(q)$), which suggests the development of the locally preferred order with decreasing $T$ and provides evidence that the model here has only weak frustration. The negligible effect of attractions on the static pair correlations indicates that the large difference seen in the relaxation dynamics cannot be explained at the static pair level~\cite{Attract3}. In fact, it has been shown that considering locally preferred structures and higher-order static correlations can help to rationalize the effect of attractions on the relaxation dynamics~\cite{Attraction1, Attraction2}. In the following, we will show that even if we consider the locally preferred bond orientational order in our models, it brings little help for understanding the large difference of the dynamics.

\subsection{Dynamic heterogeneity, bond orientational order and correlation length scales}

\begin{figure}[tb]
 \centering
 \includegraphics[angle=0,width=0.45\textwidth]{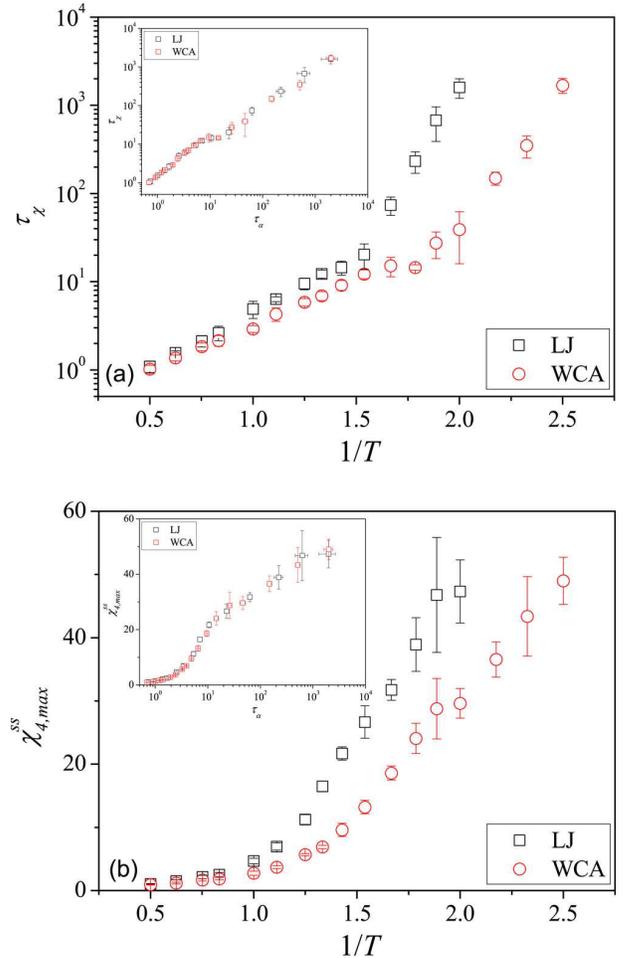}
 \caption{Arrhenius plot of (a) $\tau_{\chi}$ and (b)$\chi_{4, max}^{ss}$ for LJ and WCA models. Insets: $\tau_{\chi}$ and $\chi_{4, max}^{ss}$ as a function of $\tau_{\alpha}$ for LJ and WCA models. Note that dependence of $\tau_{\chi}$ and $\chi_{4, max}^{ss}$ on $\tau_{\alpha}$ is almost the same for the two models.}
\end{figure}

It is widely observed that the dynamic slowing down accompanies growing dynamic heterogeneity in glass-forming liquids. Therefore, it can be expected that similar effect of attractions on relaxation dynamics will occur on dynamic heterogeneity. Here, we quantify the dynamic heterogeneity by the self part of the four-point density correlations, which dominates results of the total four-point density correlations~\cite{Glotzer1, Glotzer2}. First, a time-dependent self-overlap order parameter $Q_{s}(t)$ is defined as
\begin{equation}
Q_{s}(t)=\frac{1}{N}<\sum_{j=1}^{N}w(|\textbf{r}_{j}(t)-\textbf{r}_{j}(0)|)>,
\end{equation}
with $w=1(0)$ for$|\textbf{r}_{j}(t)-\textbf{r}_{j}(0)|\leq(>)a$, where $a=0.3$ is a threshold value and choosing other appropriate values dose not change the qualitative results. The mean square variance of $Q_{s}(t)$ defines the four-point susceptibility
\begin{equation}
\chi_{4}^{ss}(t)=\frac{L^2}{N^2}[<Q_{s}(t)^{2}>-<Q_{s}(t)>^{2}],
\end{equation}
which measures the degree of the cooperativity of structural relaxation. For a typical glass-forming liquid, the peak height $\chi_{4, max}^{ss}$ of $\chi_{4}^{ss}(t)$ increases and the peak time $\tau_{\chi}$ shifts to larger times as $T$ decreases~\cite{Berthier2}. We present the results of $\tau_{\chi}$ and $\chi_{4, max}^{ss}$ for the two models in Fig. 3. We find that both models exhibit growing dynamic heterogeneities as $T$ is lowered. Attractions do lead to enhancement of dynamic heterogeneity of supercooled liquids, which has been pointed out in the previous work~\cite{Attract1, Attract2, Weeks, Chandler}. Moreover, attractions affect $\tau_{\chi}$ and $\chi_{4, max}^{ss}$ in a similar manner as compared to their influence on $\tau_{\alpha}$, which can be confirmed by plotting $\tau_{\chi}$ and $\chi_{4, max}^{ss}$ as a function of $\tau_{\alpha}$, as illustrated in the insets of Fig. 3. We observe that dependence of both $\tau_{\chi}$ and $\chi_{4, max}^{ss}$ on $\tau_{\alpha}$ is almost the same for LJ and WCA models.

\begin{figure}[tb]
 \centering
 \includegraphics[angle=0,width=0.45\textwidth]{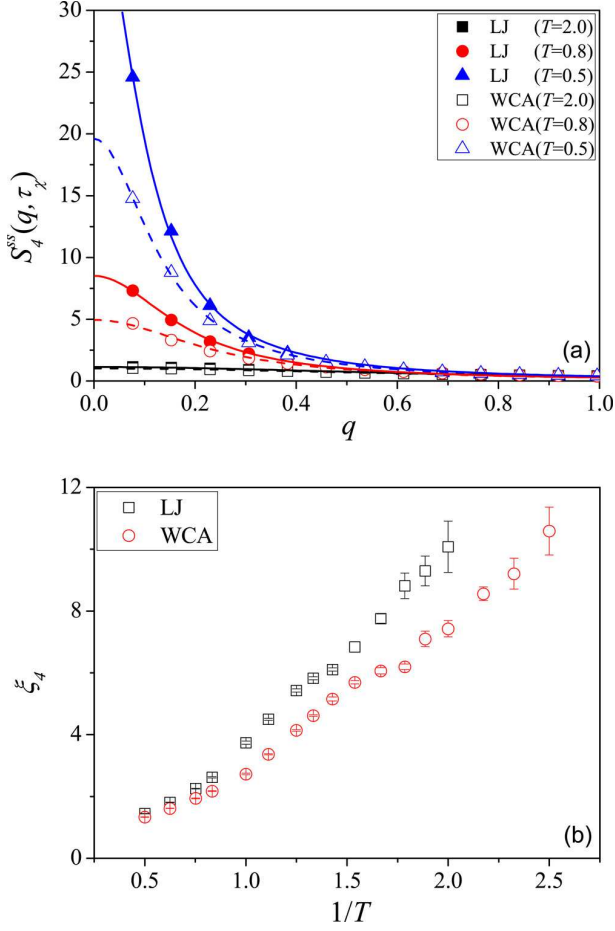}
 \caption{(a) $S_{4}^{ss}(q,\tau_{\chi})$ for LJ and WCA models at several temperatures. Lines are the results of the OZ fittings (see text). (b) Arrhenius plot of $\xi_{4}$ for LJ and WCA models.}
\end{figure}

The dynamic correlation length is investigated by the four-point, time-dependent structure factor for the self-overlapping particles, which is defined as
\begin{equation}
S_{4}^{ss}(q,t)=\frac{L^2}{N^2}<\widetilde{\rho}(q, t)\widetilde{\rho}(-q, t)>,
\end{equation}
where $\widetilde{\rho}(q,t)=\sum_{j=1}^{N}w(|\textbf{r}_{j}(t)-\textbf{r}_{j}(0)|)\exp[i\textbf{q}\cdot\textbf{r}_{j}(0)]$. Here, the time $t$ is taken as $\tau_{\chi}$, where dynamic heterogeneity becomes most pronounced~\cite{Glotzer1}. Representative results for the two models are given in Fig. 4(a). $S_{4}^{ss}(q,\tau_{\chi})$ at low-$q$ region can be well fitted by the Ornstein-Zernike (OZ) function $S_{0}/[1+(q\xi_{4})^{2}]$ and then the dynamic correlation length $\xi_{4}$ is obtained in this way. The results of $\xi_{4}$ for the two models are shown in Fig. 4(b). It is seen that difference of $\xi_{4}$ between the two models is very small at high temperatures and becomes pronounced as $T$ decreases, indicating that the same effect of attractions on relaxation dynamics and dynamic heterogeneity occurs on the dynamic correlation length. This is not surprising since $\xi_{4}$ provides similar information as $\chi_{4, max}^{ss}$ does. Therefore, attractions can strongly affect dynamic heterogeneity of glass-forming liquids.

\begin{figure}[tb]
 \centering
 \includegraphics[angle=0,width=0.45\textwidth]{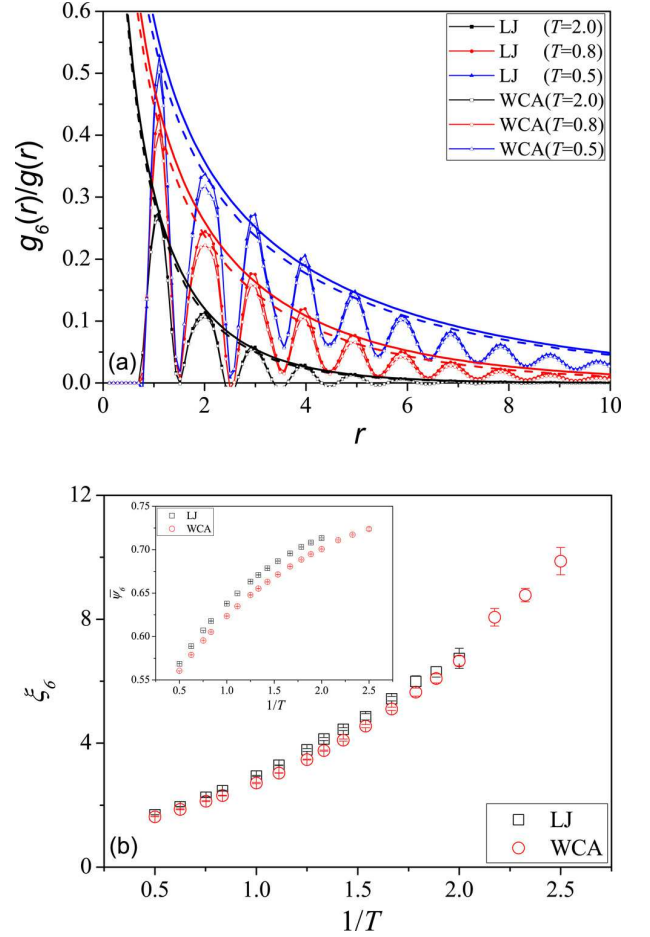}
 \caption{(a) $g_{6}(r)/g(r)$ for LJ and WCA models at several temperatures. Lines are the results of the OZ fittings (see text). (b) Arrhenius plot of $\xi_{6}$ for LJ and WCA models. Inset: the average hexagonal order parameter $\overline{\Psi}_{6}$ as a function of $1/T$.}
\end{figure}

Turning to the static properties, one may wonder whether similar effect of attractions will occur on the bond orientational order and the associated static correlation length. Since the local preferred order in our model is hexagonal, we use a sixfold bond-orientation order parameter for each particle to characterize the local structure. First, we define $\psi_{6}^{j}=\frac{1}{n_{j}}\sum_{m=1}^{n_{j}}\exp(i6\theta_{m}^{j})$ and $\Psi_{6}^{j}=|\psi_{6}^{j}|$. Here, $n_{j}$ is the number of the nearest neighbors for particle $j$ which is determined by the Voronoi construction~\cite{Allen}, and $\theta_{m}^{j}$ is the angle between $(\textbf{r}_{m}-\textbf{r}_{j})$ and the $x$ axis (particle $m$ is a neighbor of particle $j$). The order parameter of the system is defined as $\overline{\Psi}_{6}=\frac{1}{N}<\sum_{j=1}^{N}\Psi_{6}^{j}>$. The spatial correlation of $\psi_{6}^{j}$ is then calculated as
\begin{equation}
g_{6}(r)=\frac{L^2}{2\pi r\Delta rN(N-1)}<\Sigma_{j\neq k}\delta(r-|\textbf{r}_{jk}|)\psi_{6}^{j}\psi_{6}^{k*}>.
\end{equation}
The results of $g_{6}(r)/g(r)$ for several temperatures are presented in Fig. 5(a). The static correlation length $\xi_{6}$ can be obtained by fitting the envelops of $g_{6}(r)/g(r)$ to an OZ formula $r^{-1/2}\exp(-r/\xi_{6})$. We show results of $\xi_{6}$ in Fig. 5(b). In the inset of Fig. 5(b), we also present $\overline{\Psi}_{6}$ as a function of $1/T$. In both systems, we find that $\xi_{6}$ and $\overline{\Psi}_{6}$ monotonically increase in the studied $T$ range as $T$ decreases, as found in other weakly glass-forming liquids~\cite{Tanaka1, Tanaka2}. It is seen that attractions do lead to an increase of $\xi_{6}$ and $\overline{\Psi}_{6}$ in the studied $T$ range. Yet, the manner how attractions affect bond orientational order and the associated static correlation length is very different as compared to their effect on relaxation dynamics and dynamic heterogeneity, i.e., the difference of $\xi_{6}$ and $\overline{\Psi}_{6}$ between the two models is already seen in the high $T$ regime, but it does not increase with decreasing $T$ and remains nearly constant in the studied $T$ range. We note that this is reminiscent of effect of attractions on the thermodynamics of liquids, e.g., the pressure of the WCA model is roughly shifted up by a constant from that of the LJ model at fixed density~\cite{Attract2}. Obviously, the growth of bond orientational order and the static correlation length scale on the approach to the glass transition point, which might be merely a manifestation of favoring the configurational entropy in weakly frustrated glass-forming liquids, cannot rationalize the effect of attractions on relaxation dynamics.

\begin{figure}[tb]
 \centering
 \includegraphics[angle=0,width=0.45\textwidth]{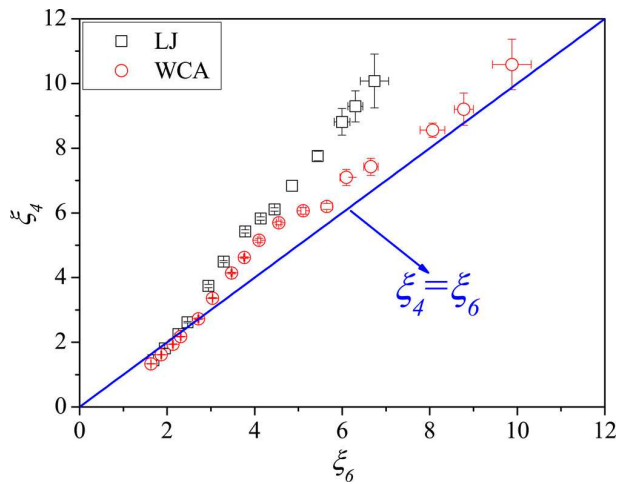}
 \caption{$\xi_{4}$ as a function of $\xi_{6}$ for LJ and WCA models. A decoupling between static and dynamic length scales is apparent, i.e., $\xi_{4}$ can be much larger than $\xi_{6}$ in the LJ model, while the difference between the two length scales is small in the WCA model.}
\end{figure}

\begin{figure}[tb]
 \centering
 \includegraphics[angle=0,width=0.40\textwidth]{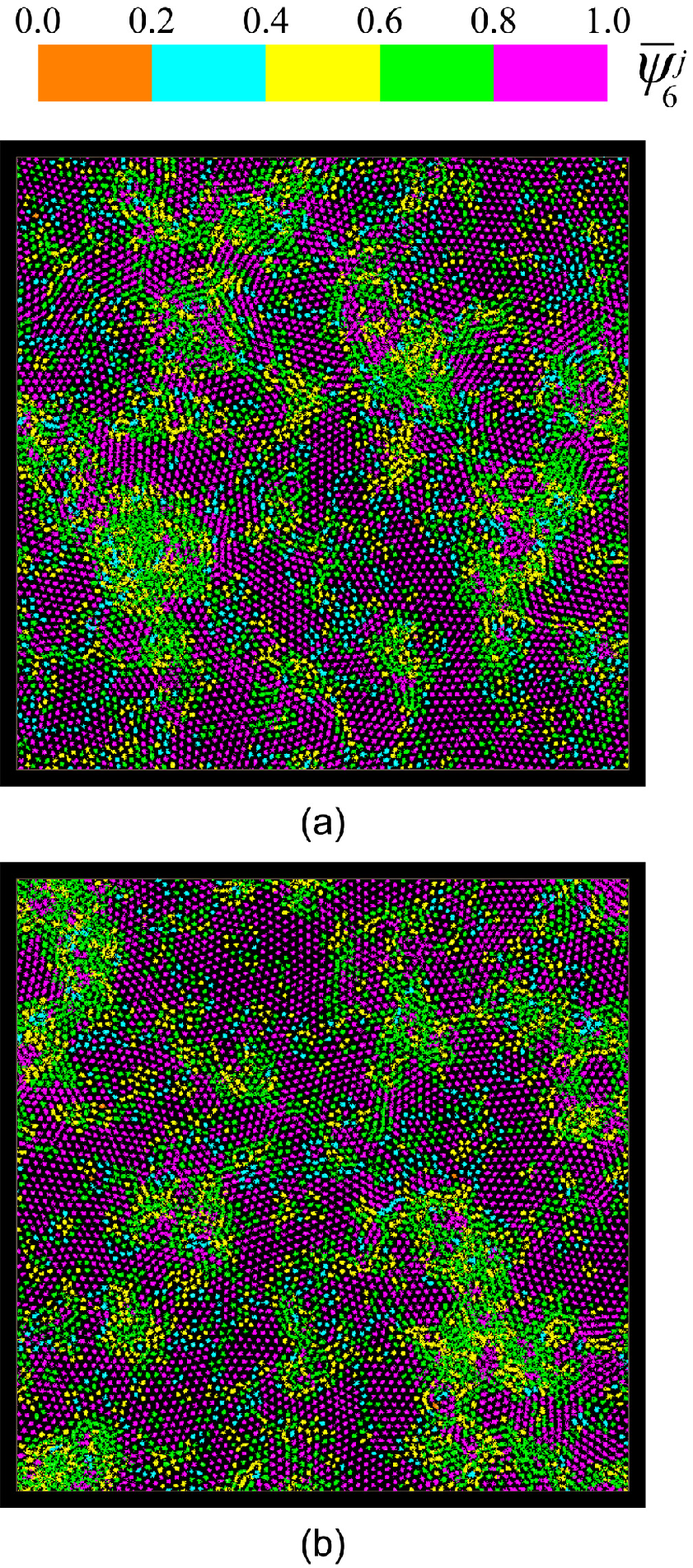}
 \caption{Trajectories of particles with different hexagonal order during an interval of $\tau_{\alpha}$ for (a) LJ model at $T=0.5$ and (b) WCA model at $T=0.4$. Note that cyan particles relax more slowly than green and yellow particles in both models, revealing an unexpected relationship between hexagonal structure and dynamic heterogeneity.}
\end{figure}

\begin{figure}[tb]
 \centering
 \includegraphics[angle=0,width=0.45\textwidth]{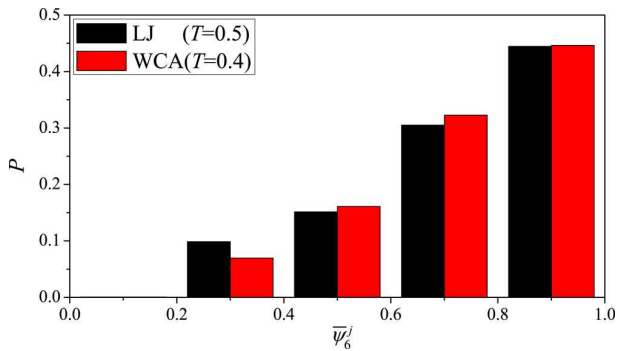}
 \caption{Probability distribution of $\overline{\Psi}_{6}^{j}$ averaged over an interval of $\tau_{\alpha}$ for LJ model at $T=0.5$ and WCA model at $T=0.4$.}
\end{figure}

The above results clearly indicate that attractions can affect dynamic heterogeneity and bond orientational order in different manners. In Fig. 6, we plot $\xi_{4}$ as a function of $\xi_{6}$ for both models, in order to make direct comparisons between dynamic and static correlation lengths and thus better illustrate the decoupling between them. It should be mentioned that the values of $\xi_{4}$ and and $\xi_{6}$ are slightly dependent on the used parameter (e.g., the threshold value $a$ in the definition of the four-point density correlations) and the data range in the fitting procedure (in this work, we obtain $\xi_{4}$ by fitting $S_{4}^{ss}(q,\tau_{\chi})$ at $q<1.0$ and $\xi_{6}$ by fitting $g_{6}(r)/g(r)$ at $r<6.0$.). However, we can still fairly compare $\xi_{4}$ and $\xi_{6}$ for the two models by employing the same criterion. It is seen in Fig. 6 that an apparent decoupling between dynamic length and static one appears, i.e., $\xi_{4}$ can be much larger than $\xi_{6}$ in the LJ model, while the difference between the two length scales is small in the WCA model. To demonstrate more clearly what order is present in the system and how the two kinds of correlation length scales differ, we monitor the relaxation process of particles with different hexagonal order for the LJ model at $T=0.5$ and the WCA model at $T=0.4$. As the dynamic correlation length is similar but the static one shows a relatively large difference at these two state points (see Figs. 3-5), we can detect how different kinds of structure relate to the growing dynamic heterogeneity. As can be seen in Fig.7, the particles with high hexagonal order (magenta dots) do dominate the relaxation process. However, the particle trajectories also reveal an unexpected relationship between hexagonal structure and dynamic heterogeneity: the particles with low hexagonal order (cyan dots) relax more slowly and thus contribute to the growing dynamic heterogeneity more than those with moderate hexagonal order (green and yellow dots). This is consistent with a previous work and the particles with low hexagonal order indeed have five or seven neighbors~\cite{Xu}. Thus, even if the average hexagonal order in the LJ model at $T=0.5$ is lower than that in the WCA model at $T=0.4$ (see the inset of Fig. 5), dynamic heterogeneity and dynamic correlation length can be similar for the two state points due to the larger amount of particles with low hexagonal order in the LJ system, as evidenced in Fig. 8, where we present the probability distribution of $\overline{\Psi}_{6}^{j}=1/\tau_{\alpha}\int_{0}^{\tau_{\alpha}}\Psi_{6}^{j}dt$ for the corresponding two state points. On the other hand, since $\xi_{6}$ indeed measures the extent to hexagonal order in a system, it can be expected that the static correlation length for the LJ model at $T=0.5$ is smaller than that for the WCA model at $T=0.4$.

\begin{figure}[tb]
 \centering
 \includegraphics[angle=0,width=0.45\textwidth]{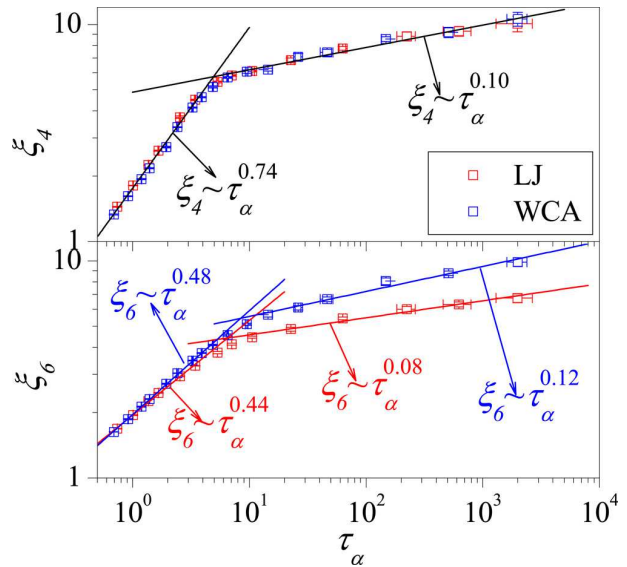}
 \caption{Log-log plot of $\xi_{4}$ (upper panel) and $\xi_{6}$ (lower panel) versus $\tau_{\alpha}$. Solid lines are power-law fits.}
\end{figure}

The different effect of attractions on dynamic heterogeneity and bond orientational order implies that bond orientational order does not provide a complete description of dynamic heterogeneity, and this conclusion can be more evident in the log-log plot of $\xi_{4}$ and $\xi_{6}$ versus $\tau_{\alpha}$, as illustrated in Fig. 9. As can be seen in the upper panel of Fig. 9, the dependence of $\xi_{4}$ on $\tau_{\alpha}$ is almost the same for LJ and WCA models and follows power-law relations (the exponent of the power law decreases from $\sim0.74$ in the high temperature regime to $\sim0.10$ in the supercooled regime), indicating that slow dynamics always accompanies growing dynamic length scales irrespective of attractions. Here, we point out that data collapse in LJ and WCA systems is also found for point-to-set correlation length and structural relaxation time~\cite{PTS4}. The situation changes for $\xi_{6}$: although we can identify power-law relations between $\xi_{6}$ and $\tau_{\alpha}$ for LJ and WCA models, the two sets of data deviate from each other and $\xi_{6}$ grows with $\tau_{\alpha}$ at a larger rate in the WCA model than that in the LJ model. It should be noted that the power-law relation between correlation length and structural relaxation time is consistent with Refs.~\cite{Glotzer1, Glotzer2, Xu} but disagrees with Refs.~\cite{Tanaka1, Tanaka2}. Therefore, even if there exists a link between static correlation length and structural relaxation time in the weakly frustrated glass-forming liquids, the bond orientational order cannot provide a complete description of dynamic heterogeneity.

\section{Conclusions}

In summary, by assessing the effect of attractions on dynamic heterogeneity, bond orientational order and the associated correlation length scales in a weakly frustrated two-dimensional glass-forming liquid, we have made a crucial examination of the role of the bond orientational order in such systems. We found that attractions affect dynamic heterogeneity and the associated dynamic length of the liquid in a similar manner as compared to their effect on relaxation dynamics. However, their influence on bond orientational order and the associated static length shows a different manner and is reminiscent of the effect of attractions on the thermodynamics of liquids. In the supercooled regime, dependence of the dynamic length on the structural relaxation time is nearly the same and shows a power-law relation for models with and without attractions. Although a power-law relation between the static correlation length and the structural relaxation time can be also identified in the same temperature range, the exponent of the power laws for the two models is not the same. This implies that the growth of bond orientational order and static correlation length scale might be merely a manifestation of favoring the configurational entropy in weakly frustrated glass-forming liquids. Our results clarify the role of bond orientational order and lead strong evidence that bond orientational order cannot provide a complete description of dynamic heterogeneity even in weakly frustrated glass-forming systems.

\begin{acknowledgments}
This work is subsidized by the National Basic Research Program of China (973 Program, 2012CB821500), and supported by the National Natural Science Foundation of China (21074137, 21222407, 50930001) programs and the fund for Creative Research Groups (50921062).
\end{acknowledgments}


\end{document}